\documentclass[12pt]{iopart}

\usepackage{graphicx}
\usepackage{amssymb}
\usepackage{braket}

\begin{document}
\title[Fabrication of high quality sub-$\mu$m JJs]{Aluminum Hard Mask Technique for the Fabrication of High-Quality Submicron Nb/Al-AlO$_x$/Nb Josephson Junctions}
\author{Ch~Kaiser$^{1}$, J~M~Meckbach$^{1}$, K~S~Ilin$^{1}$, J~Lisenfeld$^{2}$, R~Sch\"afer$^{3,4}$, A~V~Ustinov$^{2,4}$ and M~Siegel$^{1,4}$} 

\address{$^1$Institut f{\"u}r Mikro- und 
Nanoelektronische Systeme, Karlsruher Institut f\"ur Technologie (KIT), Hertzstra{\ss}e 16, D-76187 
Karlsruhe, Germany} 

\address{$^2$Physikalisches Institut, Karlsruher Institut f\"ur Technologie (KIT), Wolfgang-Gaede-Str. 1, D-76131 Karlsruhe, Germany} 

\address{$^3$Institut f\"ur Festk\"orperphysik, Karlsruher Institut f\"ur Technologie (KIT), Hermann-von-Helmholtz-Platz 1, D-76344 Eggenstein-Leopoldshafen, Germany} 

\address{$^4$Center for Functional Nanostructures, 
Karlsruher Institut f\"ur Technologie (KIT), Wolfgang-Gaede-Stra{\ss}e 1a, D-76128 Karlsruhe, Germany}

\ead{christoph.kaiser@kit.edu}

\begin{abstract}

We have developed a combined photolithography and electron-beam lithography fabrication process for sub-$\mu$m to $\mu$m-size Nb/Al-AlO$_x$/Nb Josephson junctions. In order to define the junction size and protect its top electrode during anodic oxidation, we developed and used the new concept of an aluminum hard mask. Josephson junctions of sizes down to 0.5~$\mu$m$^2$ have been fabricated and thoroughly characterized. We found that they have a very high quality, which is witnessed by the $IV$ curves with quality parameters $V_{\rm m}>50$~mV and $V_{\rm gap}=2.8$~mV at 4.2~K, as well as $I_{\rm c}R_{\rm N}$ products of $1.75-1.93$ mV obtained at lower temperatures. In order to test the usability of our fabrication process for superconducting quantum bits, we have also designed, fabricated and experimentally investigated phase qubits made of these junctions. We found a relaxation time of $T_1=26$~ns and a dephasing time of $T_2=21$~ns.

\end{abstract}


\submitto{\SUST}

\maketitle

\section{Introduction}

Nb/Al-AlO$_x$/Nb Josephson junctions (JJs) with sizes in the sub-$\mu$m to $\mu$m range are required or advantageous for many applications such as high-speed superconducting digital circuits (RSFQ), mm-wave receivers and submillimeter wave mixers, programmable voltage standards, SQUIDS and superconducting qubits. Considering the multitude of possible applications, it is not surprising that several fabrication techniques for sub-$\mu$m to $\mu$m-size junctions have been developed over the years. While a precise and reproducible definition of the JJ size is routinely performed with the help of electron-beam lithography, the application of anodic oxidation in such processes remains a challenge. This is due to the fact that the Nb$_2$O$_5$ layer formed during the anodic oxidation creeps under the resist mask protecting the top electrode of the JJ (this is known as "encroachment" \cite{subum_RHEA_1991,subum_IPHT_1999}). Even for a plasma hardened resist and an anodization voltage of 20~V, a significant encroachment was still observed \cite{subum_RHEA_1991}. Some groups solve this problem by simply omitting the anodic oxidation or performing it at very low voltages below 10~V \cite{subum_IBM_1991,subum_PTB_spinonglass_1999,subum_IRAM_2001,subum_StonyBrook_2004}. Since we found that a certain Nb$_2$O$_5$ thickness is needed to ensure low subgap leakage currents, this was not an option for us. Other groups have replaced the resist mask with an insulating hard mask such as SiO or SiO$_2$ \cite{subum_IPHT_1999,subum_Imamura_1988,subum_StonyBrook_1995,subum_PTB_anodization_1999,subum_PTB_2005}, which entirely eliminates the encroachment. These hard masks have to be removed by an additional process step, either by reactive ion etching (RIE) through the insulator right above the JJ or by chemical mechanical polishing (CMP). However, it has been found that plasma processes \cite{Topylgo_plasma_induced_damage} as well as CMP \cite{Yamamori_CMP_damage} can damage the Josephson junctions. Consequently, we have developed a novel method using an aluminum hard mask, which can be removed by a wet etching process with KOH, which should not damage the Josephson junctions. Additionally, wet etching is a very fast processing step, which helps to reduce the turnaround time.

This article is organized as follows: First, the newly developed process is discussed in some detail. Second, we present characteristic measurements, which show the high quality of our Josephson junctions for all discussed applications. Finally, we present measurements on phase qubits and evaluate the experimentally observed coherence times, in order to see if our process is suitable for the fabrication of quantum devices.

\section{Fabrication Process}

\begin{figure}[t]
	\centering
	\includegraphics[width=0.5\linewidth,angle=0,clip]{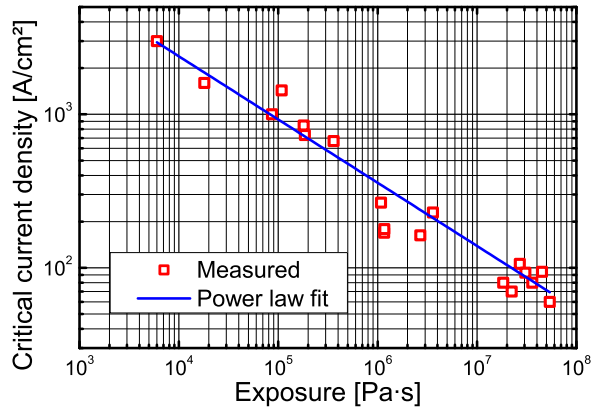}
	\caption{Critical current density of our JJs as a function of the oxygen exposure during the formation of the tunneling barrier.\label{figure_trilayer_exposure}}
\end{figure}

As a first step in the fabrication, Nb/Al-AlO$_x$/Nb trilayers are deposited {\it in-situ} on an oxidized silicon substrate in a vacuum system containing Nb and Al magnetron sputtering targets. Both Nb films are 100~nm thick and are DC sputtered at a power of $P = 300$~W in the main chamber of our two-chamber vacuum system with a base pressure of around $10^{-7}$~mbar. Special care has been taken to achieve stress-free Nb films, which is crucial for high quality JJs \cite{Imamura}. This was done by systematically varying the Ar working pressure $p_{\rm Ar}$ for a Nb film thickness of $200$ nm. Subsequently, the curvature of the silicon substrate bending under the film stress was measured, so that the latter could be determined quantitatively \cite{Singer}. For the optimal working point at $p_{\rm Ar}=7.2$~mTorr, residual resistance ratios (RRR) of 3.1 and transition temperatures of $T_{\rm c}=9.2$~K were obtained on 100~nm thick films, confirming the expected high quality of the Nb electrodes. The tunneling barrier is formed by DC sputtering of a 7 nm thick Al layer at $P=100$~W and a subsequent oxidation of this layer in the load lock of the deposition system. This oxidation is performed at room temperature in pure oxygen for $30$ min. The critical current density of the Josephson junctions can be reliably controlled in a wide range by the O$_2$ pressure during oxidation (see Figure~\ref{figure_trilayer_exposure}), so that the required parameters can be achieved for a multitude of applications, ranging from superconducting qubits with $j_{\rm c}<100$~A/cm$^2$ to RSFQ circuits at around 3~kA/cm$^2$. Afterwards, the shape of the bottom electrode (including the alignment marks for the following electron-beam lithography) is defined by positive photolithography. After this, the trilayer is patterned by subsequent reactive-ion-etching (CF$_4$ + O$_2$) of the top Nb layer, Ar ion beam etching (IBE) of the AlO$_x$ barrier and again RIE of the bottom Nb layer.

\begin{figure}[t]
	\centering
	\includegraphics[width=0.7\linewidth,angle=0,clip]{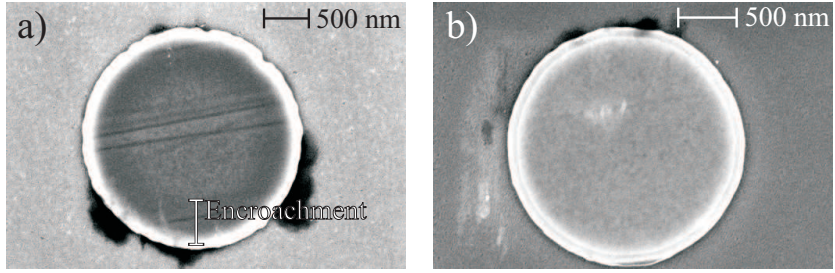}
	\caption{a) SEM pictures of Josephson junctions defined with the negative e-beam resist. The encroachement is clearly visible. b) JJ defined with the hard mask technique after removal of the Al layer. The Nb grains can be seen up to the edges of the JJ and no encroachment is observed.\label{figure_comp_masks}}
\end{figure}

In the following, the Josephson junctions are defined with the use of a hard mask. As already discussed, we wanted to avoid any further processes which might damage the junctions such as plasma etching or CMP. Hence, insulators like SiO or SiO$_2$ were ruled out as hard mask material. To find an alternative, we searched for a metal that could be anodized, so that it would not act as a shortcut for the anodization current, and that could be removed fast and selectively by a clean wet etching process. Since aluminum fulfills all of these criteria, the hard mask is created by positive electron-beam lithography and DC sputtering of a 50 nm thick Al layer with subsequent lift-off. It is then used for patterning of the junctions, which begins with etching the top electrode of the trilayer by RIE. Since aluminum is not etched here, both the AlO$_x$ tunneling barrier and the Al mask act as ideal etch stoppers. Now, the first insulating layer is formed by anodic oxidation of the samples. Here, it is taken care that all sidewalls of the lower trilayer electrode are oxidized. The anodization takes place in an aqueous solution of (NH$_4$)B$_5$O$_8$ and C$_2$H$_6$O$_2$ at room temperature. Per applied volt, 0.88 nm Nb are turned into 2.3 nm Nb$_2$O$_5$ and 0.9 nm Al are turned into 1.3 nm AlO$_x$ \cite{AO_Berkeley_2003}, so that for a typical anodization voltage of 25 V, only half the thickness of the Al hard mask is oxidized and the underlying Nb is protected. The hard mask is removed by wet etching with KOH, which is a fast process and should not damage the Josephson junctions. The JJ surface after this processing step was investigated with scanning electron microscopy (SEM), as can be seen in Figure~\ref{figure_comp_masks}. While encroachment of the Nb$_2$O$_5$ was clearly visible on JJs which were defined using electron beam resist, we found indeed no encroachment for JJs defined by the Al hard mask technique. 

\begin{figure}[b]
	\centering
	\includegraphics[width=0.7\linewidth,angle=0,clip]{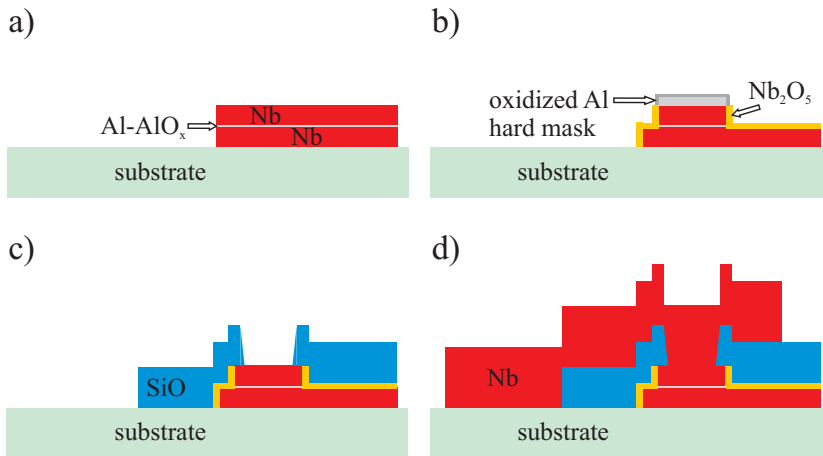}
	\caption{Schematic overview of the JJ fabrication process. a) After deposition and patterning of the trilayer. b) After JJ definition and anodic oxidation. c) After removal of the hard mask and deposition of the SiO insulation layer. d) Final structure after definition of the wiring layer.\label{figure_fabr_subum}}
\end{figure}

After that, the second insulator SiO is deposited employing positive e-beam lithography, thermal evaporation and lift-off. During evaporation, the wafers are water cooled in order to avoid thermally induced damage to the resist. Finally, the Nb wiring layer is defined by negative photolithography and deposited in a magnetron sputtering system. Here, an {\it in-situ} pre-cleaning by low-energy Ar IBE is carried out before the sputtering, in order to provide a good electrical contact between the top junction electrode and the wiring layer. This is especially required for JJs sizes in the sub-$\mu$m range where the aspect ratios of the vias through the SiO layer come close to one. An overview of the entire fabrication process can be found in Figure~\ref{figure_fabr_subum}.

\section{JJ Characterization}

\begin{figure}[t]
	\centering
	\includegraphics[width=0.5\linewidth,angle=0,clip]{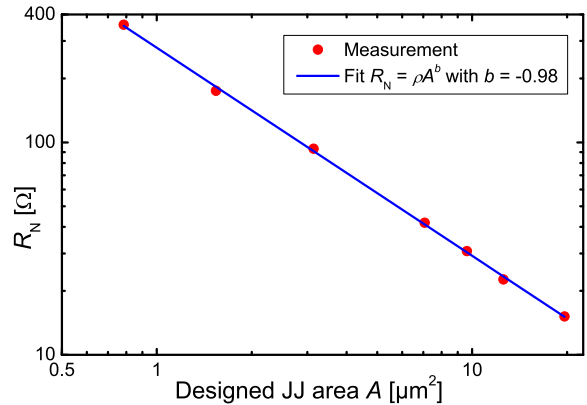}
	\caption{Normal resistance $R_{\rm N}$ over designed JJ area $A$ for various circular junctions having a critical current density of $j_{\rm c}= 660$~A/cm$^2$, measured at 4.2 Kelvin. The fit shows an excellent agreement with the expected behavior $R_{\rm N}\propto1/A$.\label{figure_JJ_size_RN}}
\end{figure}

We have fabricated junctions of various critical current densities $j_{\rm c}$ and characterized them at 4.2 Kelvin. For all applications discussed above, one of the most important quality parameters is the ratio of subgap resistance $R_{\rm sg}$ (evaluated at a voltage of 2 mV) over normal resistance $R_{\rm N}$. For the entire range of critical current densities shown in Figure~\ref{figure_trilayer_exposure}, this $R_{\rm sg}/R_{\rm N}$ ratio was typically larger than 30, which indicates a very high quality. In the following, the further discussion of junction quality will be carried out on two exemplary JJ batches having different critical current density values.

\begin{figure}[t]
	\centering
	\includegraphics[width=0.5\linewidth,angle=0,clip]{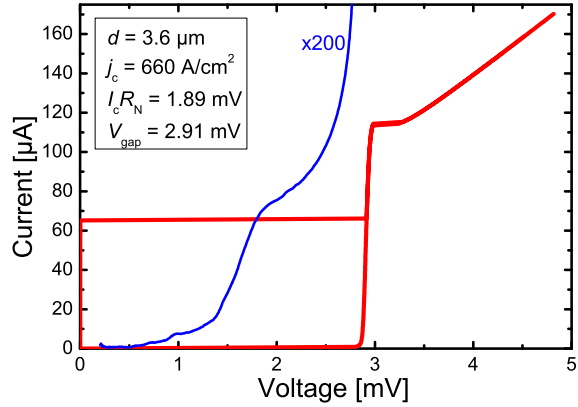}
	\caption{$IV$ curve of a JJ with $j_{\rm c}= 660$~A/cm$^2$ and a diameter of $d=3.6$~$\mu$m, measured at $T=20$~mK. High quality parameters $I_{\rm c}R_{\rm N}=1.89$~mV and $V_{\rm gap}=2.91$~mV are obtained. The subgap branch (magnified by a factor of 200) shows low leakage currents; it has been measured with a voltage bias and an applied magnetic field along the JJ area, suppressing the critical current to 8~\% of its zero-field value.\label{figure_IV_jj23}}
\end{figure}

For a trilayer with $j_{\rm c}=660$~A/cm$^2$, we systematically varied the JJ size between 0.8 and 20~$\mu$m$^2$ in order to see if the JJ geometries were defined precisely. Since critical currents $I_{\rm c}<10$~$\mu$A are strongly suppressed at $T=4.2$~K due to thermal noise, we took the normal resistance as the characteristic junction parameter. Figure~\ref{figure_JJ_size_RN} shows the measured $R_{\rm N}$ values versus the designed JJ areas $A$. The data were fit by the power law $R_{\rm N}=\rho\cdot A^b$. It was found that $b=-0.98$, so that an excellent agreement with the expected behavior $R_{\rm N}\propto1/A$ was obtained, which shows that the JJ sizes are accurately defined by the Al hard mask process. This was confirmed by occasional measurements of the JJ size with scanning electron microscopy (SEM). Furthermore, the fit value of $\rho=279$~$\Omega\mu$m$^2$ leads to an average value of $I_{\rm c}R_{\rm N}=j_{\rm c}\rho=1.84$~mV. These high $I_{\rm c}R_{\rm N}$ products show that we have strong Cooper pair tunneling close to the expected Ambegaokar-Baratoff behavior \cite{Ambegokar-Baratoff}. Four of these samples were cooled down to $T=20$~mK, so that measurements with unsuppressed critical currents were possible; an exemplary $IV$ curve can be seen in Figure~\ref{figure_IV_jj23}. Here, we observed no excess currents and $I_{\rm c}R_{\rm N}$ products of $1.75-1.93$~mV, which confirms the values given above. Furthermore, the gap voltages were found to be in the range of $V_{\rm gap}=2.88-2.92$~mV, which indicates a high quality of the Nb electrodes. The subgap branches were measured with a voltage bias and showed very low leakage currents, as can be seen in Figure~\ref{figure_IV_jj23}.

\begin{figure}[b]
	\centering
	\includegraphics[width=0.5\linewidth,angle=0,clip]{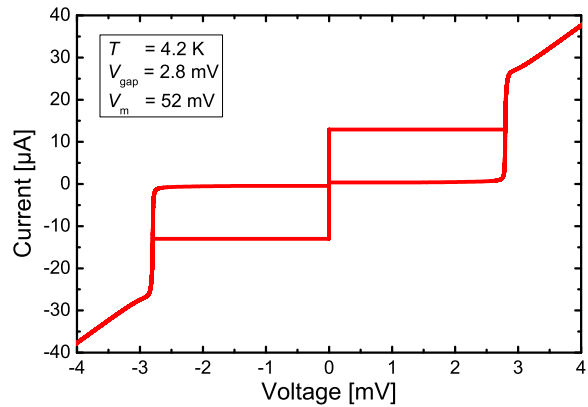}
	\caption{$IV$ curve of a circular junction with an area of 0.5$\,\mu$m$^2$ and a critical current density of $j_{\rm c}= 3\,$kA/cm$^2$, measured at 4.2 Kelvin.\label{figure_IV_JC46_JJ24}}
\end{figure}

A different trilayer with $j_{\rm c}= 3$~kA/cm$^2$ (this $j_{\rm c}$ value is of particular interest for the realization of RSFQ circuits \cite{RSFQ3kA}) led to junctions with higher $I_{\rm c}$ values. Although the critical currents in the range of $10 - 20$~$\mu$A were still somewhat suppressed by thermal noise at $T=4.2$~K, we could now perform a full characterization at this temperature. The recorded gap voltages accounted for $\geq 2.8$~mV and the characteristic voltages $V_{\rm m}=I_{\rm c}R_{\rm sg}$ were all larger than 50~mV. For JJs with areas $>1$~$\mu$m$^2$, we observed values $V_{\rm m}\geq 60\,$mV, which we contribute to a less suppressed $I_{\rm c}$. The $IV$ curve of a sample with an area of 0.5~$\mu$m$^2$ can be seen in Figure~\ref{figure_IV_JC46_JJ24}.

Since the $R_{\rm sg}/R_{\rm N}$ ratios were very high for all critical current densities (see above) and high $V_{\rm gap}$, $I_{\rm c}R_{\rm N}$ and $V_{\rm m}$ values were found at the exemplary critical current densities $j_{\rm c}= 660$~A/cm$^2$ and $j_{\rm c}= 3$~kA/cm$^2$, we can conclude that a very high junction quality is obtained by the presented process. Furthermore, the junction size was accurately defined by the Al hard mask technique.

\section{Characterization of Phase Qubit}

\begin{figure}[b]
	\centering
	\includegraphics[width=0.8\linewidth,angle=0,clip]{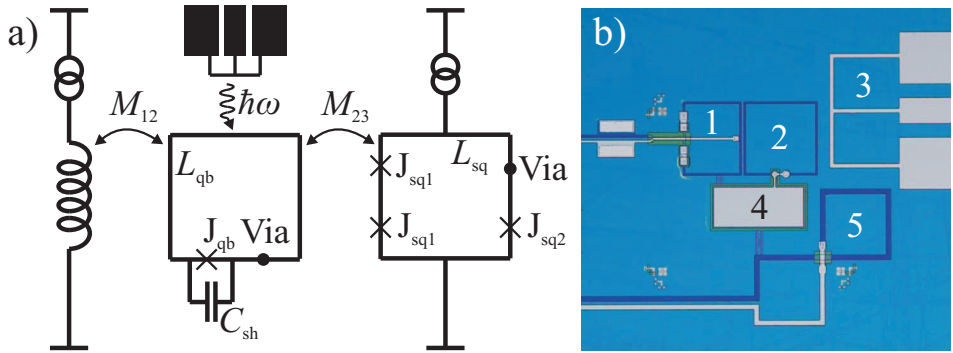}
	\caption{a) Circuit schematics of the phase qubit that was investigated. All junctions are circular and their diameters are given. The designed circuit parameters account for: $d_{\rm J,qb}=3.1\,\mu$m, $d_{\rm J,sq1}=2.83\,\mu$m, $d_{\rm J,sq2}=2\,\mu$m, $C_{\rm sh}=0.82\,$pF, $L_{\rm qb}=257\,$pH, $L_{\rm sq}=227\,$pH, $M_{12}=1.36\,$pH, $M_{23}=22\,$pH. b) Photograph of the same structure with 1 = Bias coil, 2 = Qubit, 3 = SQUID, 4 = $C_{\rm sh}$, 5 = MW line.\label{figure_phase_qubit_geometry}}
\end{figure}

High quality parameters extracted from the $IV$ curves do not necessarily mean that the junctions can be successfully used for the fabrication of superconducting quantum bits, since decoherence mechanisms at higher frequencies, which are not detected in the $IV$ curves, play an important role for such devices \cite{Martinis_TLF_coherence,McDermott}. Consequently, to see if the technological process described in this article is suitable for quantum bits, we fabricated phase qubits, of which one was characterized at mK temperatures in a dilution refrigerator. We used a critical current density of $j_{\rm c}=65$~A/cm$^2$ and a standard phase qubit design similar to the one used in \cite{Lisenfeld_Nb_vs_Al}. All vias necessary for the connection of the wiring to the lower electrode were not realized as large junctions, but the tunneling barrier was etched away with IBE, so that real vias with a direct connection between the Nb bottom electrode and the Nb wiring layer were formed. Schematics and a photograph of the circuit as well as design values can be found in Figure~\ref{figure_phase_qubit_geometry}. The design values lead to a characteristic qubit parameter of $\beta_L=2\pi L_{\rm qb}I_{\rm c,qb}/\Phi_0\simeq 3.8$. 

The setup and procedure of the measurement have been successfully used for the investigation of other phase qubits before and are described elsewhere \cite{Lisenfeld_Nb_vs_Al}. The qubit is read out by application of a short (1 ns) DC flux pulse, during which one magnetic flux quantum tunnels into the qubit loop only if the qubit was in the excited state $\ket{1}$. The associated flux signal is detected by the inductively coupled SQUID, which results in two separated SQUID switching current histograms depending on whether the qubit was in the ground state $\ket{0}$ or the excited state $\ket{1}$. These switching histograms could be used to calculate the escape probability $P_{\rm esc}$, which directly reflects the population of the qubit states. First, by measuring the distance of these two separate SQUID switching peaks and taking into account the slope of the SQUID $I_{\rm c}(\Phi)$ modulation, we determined the strength of the qubit flux signal in the SQUID to be $\Phi_{\rm sq,qb}=73$~m$\Phi_0$. This is slightly higher than the expected design value of $\Phi_{\rm sq,qb}=M_{23} \cdot I_{\rm c,qb} =53$~m$\Phi_0$. Such an increase in the qubit signal strength might be due to a higher critical current of the qubit $I_{\rm c,qb}$ than designed, which goes along well with the fact that also the observed average qubit level splitting $\Delta_{01}\approx 13$~GHz was slightly higher than expected.

\begin{figure}[t]
	\centering
	\includegraphics[width=0.5\linewidth,angle=0,clip]{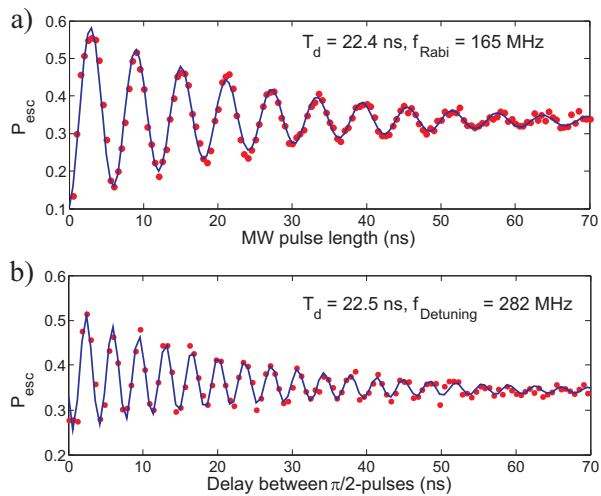}
	\caption{a) Rabi oscillations between the qubit states $\ket{0}$ and $\ket{1}$ at a level splitting of $\Delta_{01}=14.48\,$GHz. b) Ramsey oscillations for a detuning of 282 MHz from this working point.\label{figure_qubit_rabi_ramsey}}
\end{figure}

In the following, we chose a working point of $\Delta_{01}=14.48$~GHz to characterize the qubit. Rabi oscillations of the system can be seen in Figure~\ref{figure_qubit_rabi_ramsey}a. The oscillations decay exponentially with a characteristic time $T_{\rm d}=22.4\,$ns. Furthermore, the relaxation time of the qubit was found to be $T_1=26$~ns. The dephasing time $T_2$ was determined by Ramsey-type experiments with various detuning values of the applied microwave from the resonance frequency. An example of such a measurement can be seen in Figure~\ref{figure_qubit_rabi_ramsey}b. We determined the decay times of these exponentially decaying oscillations for various detuning values and took the mean value $T_2=21$~ns as our dephasing time. Details about each measurement procedure can be found in \cite{Lisenfeld_Nb_vs_Al}.

For most Nb/Al-AlO$_x$/Nb phase qubits, the relaxation times were found to be between $T_1\simeq 7$~ns and $T_1\simeq 20$~ns \cite{Lisenfeld_Nb_vs_Al,Martinis_first_current_biased_qubit,Maryland_IEEE_2007,Feofanov_FM_barrier,Maryland_Nb_vs_Al,Maryland_Multilevel_effects,Lukens_Nb_phase_qubit}. For all these devices, the dephasing times $T_2$ could either not be measured or were shorter than the $T_1$ times. This means that we have reached coherence times $T_1$ and $T_2$ which are certainly not shorter, but rather longer than for most other Nb/Al-AlO$_x$/Nb phase qubits. Consequently, we can conclude that the novel Al hard mask technique does not add any new sources of decoherence to the system and that the presented process is suited for the fabrication of quantum devices. In the future, it would be interesting to measure the coherence times of qubits having identical design but being fabricated with and without the Al hard mask technique, in order to see if this processing step has a systematic influence on decoherence.

\section{Conclusions}

We have developed a technological process for the fabrication of high quality sub-$\mu$m to $\mu$m-size Nb/Al-AlO$_x$/Nb Josephson junctions. This process employs an Al hard mask, which acts as a perfect etch stop for RIE and can be removed by a clean and fast wet etching step. Characterization of our junctions at 4.2 K and mK temperatures showed that we have high quality parameters $V_{\rm m}>50\,$mV, $V_{\rm gap}= 2.8$~mV at $T=4.2$~K and $V_{\rm gap}= 2.88-2.92$~mV, $I_{\rm c}R_{\rm N} = 1.75 - 1.93\,$mV at $T=20$~mK. We fabricated phase qubits and characterized one of them at mK temperatures. It exhibited a relaxation time of $T_1=26\,$ns and a dephasing time of $T_2=21\,$ns, which is comparable or even longer than for most other reported Nb/AlO$_x$/Nb phase qubits.

\ack
This work was partly supported by the DFG Center for Functional Nanostructures, project number B1.5. We also thank H~J~Wermund, A~Stassen and K-H~Gutbrod for technical support.

\end{document}